\documentclass[12pt,preprint]{aastex}
\slugcomment{Submitted on \today}

\shortauthors{Yi et al.}
\shorttitle{The $Y^2$ Stellar Evolutionary Tracks}

\begin{document}

\title{The Y$^2$ Stellar Evolutionary Tracks}

\author{Sukyoung K. Yi}
\affil{University of Oxford, Astrophysics, Keble Road, Oxford OX1 3RH, UK
\\ yi@astro.ox.ac.uk}

\author{Yong~-Cheol Kim}
\affil{Department of Astronomy, Yonsei University, Seoul 120-749, Korea \\
kim@galaxy.yonsei.ac.kr}

\and

\author{Pierre Demarque}
\affil{Yale University, Department of Astronomy, PO Box 208101, New Haven,
CT 06520-8101, USA \\ demarque@astro.yale.edu}

\begin{abstract}
 
	We present a database of the latest stellar models
of the $Y^2$ (Yonsei-Yale) collaboration.
	This database contains the stellar evolutionary tracks
from the pre-main-sequence birthline to the helium core flash that were
used to construct the $Y^2$ isochrones \citep{YI01,KIM02}.
	We also provide a simple interpolation routine that generates 
stellar tracks for given sets of parameters (metallicity, mass, and
$\alpha$-enhancement).

\end{abstract}


\section{Introduction}

	The $Y^2$ group has recently released their updated isochrones
\citep{YI01,KIM02}.
	Theses isochrones have been tested against many high-quality 
observational data and proved reasonably accurate.
	Besides such successes, their wide coverage in metallicity and age, 
which is useful to population synthesis studies, triggered many requests
for the corresponding stellar evolutionary tracks.
	Stellar tracks are the basic building blocks of isochrones, and thus
it is important to ascertain their accuracy by all means before we feel
confident about isochrones.
	A compelling test of such kinds is made on binary
data which can yield mass and radius information on individual stars
independently of the infamous uncertainty in distance.

\section{Stellar Evolution Models}

	The critical pieces of input physics and model parameters
are listed in Table 1, while greater details are available
from \citet{YI01} and from \citet{KIM02}.
	The models were evolved from the pre-main-sequence (MS) stellar 
birthline to the onset of helium burning in the core at the red giant 
branch tip (RGBT).
	The mass range is approximately 0.4 $M_\odot$ through 5 $M_\odot$,
which has been chosen for constructing full isochrones from MS to RGBT 
for ages 0.1 -- 20\,Gyr.

	We release the entire database of the $Y^{2}$ stellar models in 
the form shown in Tables 2 and 3.
	Each stellar track is stored in two files.
	Track1 files hold information simply for 24 select young ages 
(1 -- 80 Myr).
	In case of relatively low-mass models, this file represents 
a pre-MS track.
	Track2 files are for older ages extended to RGBT.
	A total of 150 grids, which some might want to call 
{\it equal evolutionary points}, are chosen according to a highly 
complicated scheme.
	The first 50 grids are based on the central helium abundance ($Y_{c}$)
which grows monotonically and gradually as soon as central hydrogen 
burning begins. 
	The next 100 grids are based on the core mass ($M_{c}$) that starts
growing about the time $Y_{c}$ stops increasing after reaching 
its maximum.
	This transition may be regarded as the 
{\it terminal-age main sequence}.
	The grids in $Y_{c}$ and $M_{c}$ are not fixed but variable 
in different tracks
because their maximum values and the ways they reach their maximum values are 
complex functions of mass, metallicity and so on.
	Therefore, in addition to the usual track information 
(age, temperature, luminosity), we present $Y_{c}$ and $M_{c}$ as well.

	Figure 1 shows a sample set of stellar models.
	For clarity, pre-MS parts appear separately in dotted lines.
	For this separation, we arbitrarily chose the point where 
$Y_{c} $ is 0.2\% enhanced from the initial value, which means that the star
has just begun burning atomic hydrogen in its core.
	Note that the original tracks, as shown in this figure, have orders 
of magnitude finer evolutionary grids and thus are slightly more accurate 
but far less portable than the tracks ditributed through this paper.
	If high accuracy is required for good scientific reasons, 
the original-grid tracks can be made available for a set of original
parameters ($Z$, $M$, [$\alpha$/Fe], overshoot) upon request to the authors.

\section{Database and Interpolation Code}

	The full set of stellar models and a FORTRAN package that work
for mass, metallicity, and $\alpha$-enhancement interpolation are
available from the authors upon request or directly from our Web sites 
listed below, as well as from the electronic edition of the 
{\it Astrophysical Journal}.

\begin{itemize}
\item {\tt UK: www-astro.physics.ox.ac.uk/$\sim$yi/yystar.html}
\item {\tt US: www.astro.yale.edu/demarque/yystar.html}
\item {\tt Korea: csaweb.yonsei.ac.kr/$\sim$kim/yystar.html}
\end{itemize}

	The interpolation code adopts {\it a polynomial interpolation scheme}. 
	We encourage the readers to use it when they find our model grids 
are not sufficiently small for their purposes.
	However, it should be emphasized that the interpolated 
tracks do not reproduce exactly the genuine tracks constructed 
using the stellar evolution code, as some properties evolve in much more
complicated manners within the given grid space than a simple interpolation
can mimic.
	The interpolation near the critical mass ($M^{conv}_{crit}$), 
above which a stellar core becomes convective on the MS \citep{YI01}, 
is a good example.
	Theory suggests a transition from a radiative core
to a convective core as stellar mass reaches $M^{conv}_{crit}$, 
but determining $M^{conv}_{crit}$ is not trivial because 
the development of a convective core is gradual with respect to mass.
	In this regard, the accuracy of our critical masses 
listed in Table 2 of \citet{YI01} is no better than our typical mass 
grid, $0.1 M_{\odot}$.
	As a result, interpolated tracks based on the given models
are unavoidably subject to such internal uncertainties.
	Besides, the abrupt change in the overshoot parameter near 
$M^{conv}_{crit}$ (overshoor is numerically turned on at 
$M > M^{conv}_{crit}$) introduces, perhaps inaccurately, discontinuity in
the shape of the main-sequence turn-off region as a function of stellar
mass, which makes the interpolations near  $M^{conv}_{crit}$ 
difficult.
	Keeping these inherent limitations in mind, we purposely kept 
the interpolation scheme as simple as possible.
	We suggest that the readers should use the models and code with
proper care.

\acknowledgments
This research has been supported by Korean Research Foundation Grant
KRF-2002-070-C00045 (YCK), and received partial support from 
NASA grant NAG5-8406 (PD).

% \aap \apj \apjl \aj \pasp \mnras \apjs\zap

\begin{table*}
\caption{Input Physics and Parameters} \label{tbl-1}
\begin{center}
\begin{tabular}{cc}
\tableline
\tableline
Input Parameters		& Description\\
\tableline
Solar mixture			& \citet{GN93}\\
$\alpha$-enhancement		& \citet{Van00}: [$\alpha$/Fe]=0.0, +0.3, +0.6\\
OPAL Rosseland mean opacities	& \citet{RI95}, \citet{IR96}\\
Low temperature opacities	& \citet{AF94}\\
Equations of state		& OPAL EOS \citep{Rog96} \\
Energy generation rates		& Bahcall \& Pinsonneault (1992; 1994 priv. comm.) \\
Neutrino losses			& \citet{Ito89} \\
Convective core overshoot 	& 0.2 $H_{p}$ when convective core develops\\
Helium diffusion 		& \citet{Tho94}\\
Mixing length parameter		& ${\it l}/H_{p} = 1.7431$ \\
Primordial helium abundance 	& $Y_{0} = 0.23$ \\
Helium enrichment parameter 	& $\Delta$$Y$/$\Delta$$Z = 2.0$ \\
Mass		 		& 0.4 -- 5.0~$M_{\odot}$ with $dM \approx 0.1$\\
Total Metallicity		& $Z$ = 0.00001, 0.0001, 0.0004, 0.001, 0.004, \\
				& 0.007, 0.01, 0.02, 0.04, 0.06, 0.08 \\
\tableline
\end{tabular}
\end{center}
\end{table*}

\begin{table*}
\caption{The $Y^{2}$ stellar models: Track1 file for 24 select young ages\tablenotemark{a}.}\label{tbl-2}
\begin{center}
\begin{tabular}{cccccc}
\tableline
\tableline
 N  & Time(Gyr) & logT & log $L/L_{\odot}$  & $Y_{core}$ & $M_{core}/M_{\odot}$ \\
\tableline
  1  &  0.00100000 &  3.65407119 &  0.26487544  & 0.26997100  & 0.00000000 \\
  2  &  0.00200000 &  3.65228210 &  0.07558276  & 0.26997121  & 0.00000000 \\
  3  &  0.00300000 &  3.65216411 & -0.03393051  & 0.26997175  & 0.00000000 \\
  4  &  0.00400000 &  3.65138939 & -0.11382817  & 0.26997235  & 0.00000000 \\
  5  &  0.00500000 &  3.65210960 & -0.16933076  & 0.26997302  & 0.00000000 \\
\tableline
\end{tabular}
\tablenotetext{a}{Table 2, together with our stellar model interpolation code, 
is available in its entirety in the electronic edition of the
{\it Astrophysical Journal Supplement.} A portion is shown here for 
guidance regarding its form and content.}
\end{center}
\end{table*}

\begin{table*}
\caption{The $Y^{2}$ stellar models: Track2 file for MS through RGBT\tablenotemark{a}.}\label{tbl-3}
\begin{center}
\begin{tabular}{cccccc}
\tableline
\tableline
 N  & Time(Gyr) & logT & log $L/L_{\odot}$  & $Y_{core}$ & $M_{core}/M_{\odot}$ \\
\tableline
  1   &  0.04028355 &   3.56415623 &  -1.45710618  &  0.23199999  &  0.00000000\\
  2   &  0.06581653 &   3.57083627 &  -1.51306541  &  0.23202999  &  0.00000000\\
  3   &  0.08711833 &   3.57638109 &  -1.53573156  &  0.23205999  &  0.00000000\\
  4  &   0.10497279 &   3.57789166 &  -1.55702151  &  0.23208999  &  0.00000000\\
  5  &   0.11848279 &   3.57678274 &  -1.57652950  &  0.23211999  &  0.00000000\\
\tableline
\end{tabular}
\tablenotetext{a}{Table 3, together with our stellar model interpolation code, 
is available in its entirety in the electronic edition of the
{\it Astrophysical Journal Supplement.} A portion is shown here for 
guidance regarding its form and content.}
\end{center}
\end{table*}

\clearpage

\begin{figure}
\plotone{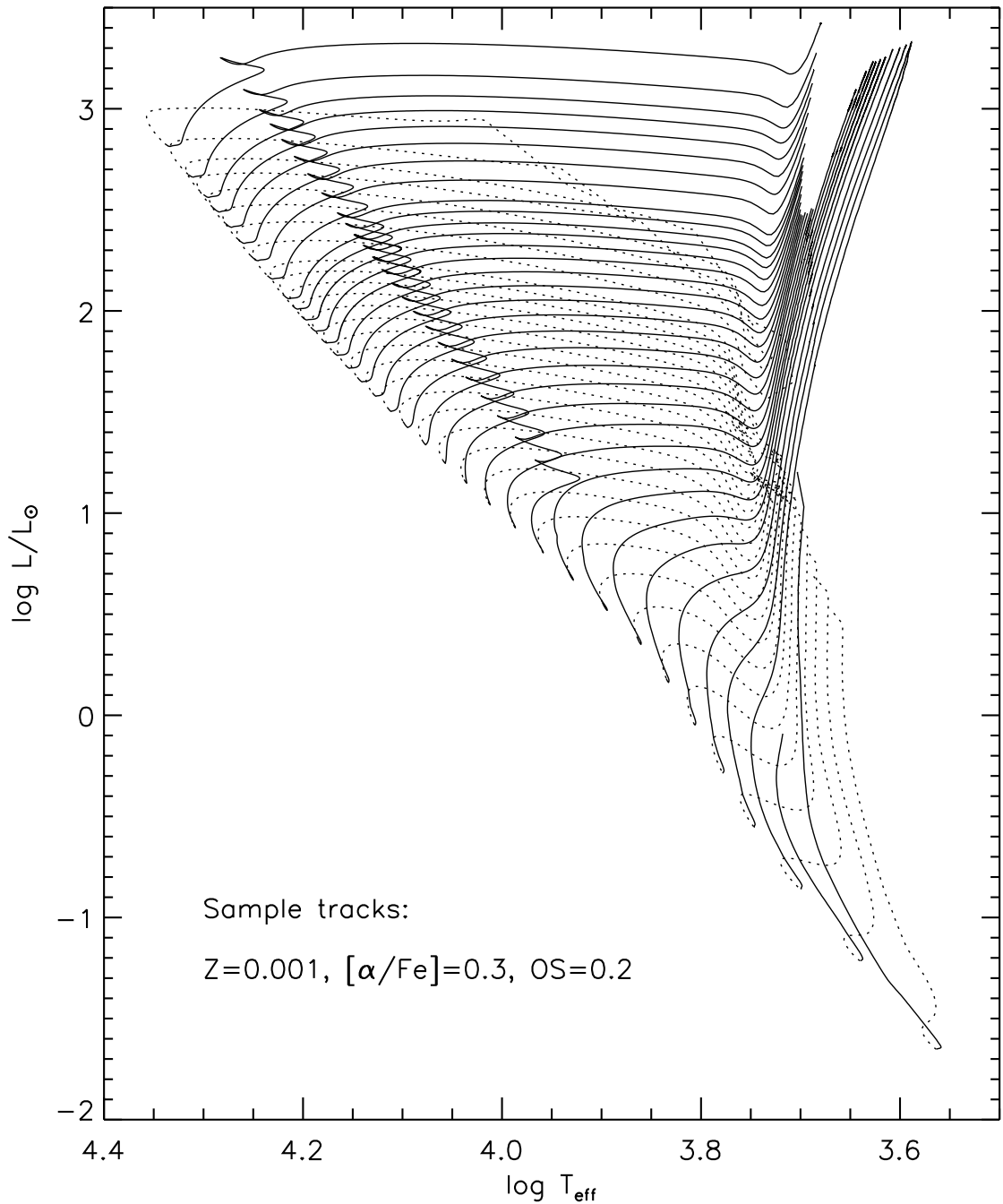}
\caption[f1.eps]{
A sample set of stellar evolutionary tracks for a mass range 0.4 (faintest) 
-- $5.0 M_{\odot}$ (brightest). Pre-MS parts are shown in dotted line while 
further evolution is shown in continuous line.
\label{f1}}
\end{figure}

\end{document}